\documentclass[twocolumn,showpacs,preprintnumbers,amsmath,amssymb,superscriptaddress, prl,floatfix]{revtex4-1}

\usepackage{graphicx}
\usepackage{dcolumn}
\usepackage{tabularx}
\usepackage{bm}
\usepackage{epstopdf}
\usepackage{amsmath}

\begin{document}

\title{Stochastic resonance in collective exciton-polariton excitations inside a GaAs microcavity}

\author{H. Abbaspour}
\affiliation{Laboratory of Quantum Optoelectronics, \'Ecole Polytechnique F\' ed\'erale de Lausanne, CH-1015, Lausanne, Switzerland}
\author{S. Trebaol}
\email[E-mail:]{stephane.trebaol@epfl.ch}
\affiliation{Laboratory of Quantum Optoelectronics, \'Ecole Polytechnique F\' ed\'erale de Lausanne, CH-1015, Lausanne, Switzerland}
\author{F. Morier-Genoud}
\affiliation{Laboratory of Quantum Optoelectronics, \'Ecole Polytechnique F\' ed\'erale de Lausanne, CH-1015, Lausanne, Switzerland}
\author{M. T. Portella-Oberli}
\affiliation{Laboratory of Quantum Optoelectronics, \'Ecole Polytechnique F\' ed\'erale de Lausanne, CH-1015, Lausanne, Switzerland}
\author{B. Deveaud}
\affiliation{Laboratory of Quantum Optoelectronics, \'Ecole Polytechnique F\' ed\'erale de Lausanne, CH-1015, Lausanne, Switzerland}

\date{\today}

\begin{abstract}
We report the first observation of stochastic resonance in confined exciton-polaritons. We evidence this phenomena by tracking the polaritons behavior through two stochastic resonance quantifiers namely the spectral magnification factor and the signal-to-noise ratio. The evolution of the stochastic resonance in function of the modulation amplitude of the periodic excitation signal is studied. Our experimental observations are well reproduced by numerical simulations performed in the framework of the Gross-Pitaevskii equation under stochastic perturbation.
\end{abstract}


\maketitle

Noise is generally considered as an harmful contribution in usual circumstances. Stochastic resonance is an intriguing effect allowing to enhance coherently the response of a nonlinear process by the addition of a noisy perturbation \cite{Badzey2005}. Synergetic interplay between a feeble input periodic signal and the stochastic perturbation allows to amplify coherently the response of a nonlinear system. Suggested in the 80's to explain the ice age occurrence \cite{Benzi1981}, stochastic resonance have been demonstrated in a wide scope of fields as chemistry, biology, physics, and medicine \cite{Gammaitoni1998,Hanggi2002,Giacomelli1999}. The first experimental demonstration was obtained in a Schmitt-Trigger circuit \cite{Fauve1983}. This pioneer work stimulates researches and among them, observations of stochastic resonance in optical bistable regime have been reported in  bistable laser \cite{McNamara1988}, passive optical cavity  \cite{Dykman1991} and atomic system in electromagnetically induced transparency configuration\cite{Joshi2006}.\\
This counterintuitive phenomena stimulated a large body of theoretical works on classical stochastic resonance in double well potential \cite{Bartussek1994}, excitable systems and also in more exotic systems as coupled processes  \cite{Wellens2004}. Quantum stochastic resonance has been also suggested theoretically \cite{Lofstedt1994}, but up to now, an experimental demonstration is still lacking.\\
Key ingredients to observe stochastic resonance are (i) any kind of threshold, (ii) a coherent periodic signal and (iii) a noisy environment. Some proposals even mention the possibility of observing stochastic resonance in a thresholdless system \cite{Bezrukov1997}. Nowadays, stochastic resonance is not only an appealing phenomena, but also a potential means for several purposes like signal and image processing \cite{Dylov2010} or neuron transmission enhancement in neurobiology \cite{Douglass1993}. Thereby, this effect can be seen as a locking technique to extract and amplify weak signal buried in noise \cite{Nishiguchi2012}.\\
Exciton-polaritons are quasiparticles resulting from the strong coupling between photons and excitons embedded in a semiconductor microcavity. Their properties, inherited from the exciton part, give rise to nonlinear effects that can be driven and read out through their photonic counterpart coupled in and out of the cavity. Moreover, polaritons carry a pseudospin that brings up spinor related nonlinear effects \cite{Takemura2014}. Polariton fluids, by their unique properties, are a model playground to study both fundamental and applied physics phenomena.\\
Since their first experimental observation about 20 years ago \cite{Weisbuch1992}, microcavity polaritons have been the matter of important demonstrations such as BEC \cite{Kasprzak2006}, superfluid effects \cite{Amo2009,Grosso2011,Kohnle2011,Kohnle2012}, bistability \cite{Baas2004}, multistability \cite{Paraiso2010} and more recently polariton based devices for signal processing \cite{Cerna2013,Amo2010,Gao2012}. As in other systems, noise is usually considered as a detrimental and unavoidable effect in polariton systems \cite{Johne2009}. Conversely, taking advantage of noise to keep and control polariton properties would be a powerful tool. Actually, this would be possible by reaching the stochastic resonance condition in polariton system. The half-light half-matter nature of polaritons opens the possibility to study random behaviors introduced by either their photon or exciton part raising them as a versatile system to implement such studies.\\ 
The recent demonstration of a spin trigger based on polariton fluid \cite{Cerna2013} can be addressed in the context of stochastic resonance. The ability to drive the system in a specific spin state by the contribution of a spin noise would be an original contribution to the domain. Progress in Molecular Beam Epitaxy, permits to tailor semiconductor microcavities energy potential landscape \cite{Kaitouni2006,Jacqmin2014,Galbiati2012,Tanese2014}. Coupled structures in semiconductor materials will allow investigating the effect of noise in an array of coupled polariton localized spots. Eventually polariton fluids seem to be a good playground for investigation of noise induced order in spatiotemporal systems \cite{Neiman1995}.\\
We report here the first demonstration of stochastic resonance in microcavity exciton-polaritons. This effect is evidenced through accurate criteria namely the spectral magnification factor and the Signal-to-Noise Ratio (SNR). The robustness of the process is witnessed as a function of the modulated amplitude of the weak coherent driving field. Experimental observations are reproduced by a numerical study based on the Gross-Pitaevskii equation driven by a stochastic excitation.\\
We demonstrate stochastic resonance in GaAs microcavity where polaritons are confined in mesa structures of 3 $\mu$m diameter \cite{Kaitouni2006}. This structure can be seen as a model system to implement such a study on stochastic resonance. Zero dimensional (0D) polaritons display discrete energy levels that can be addressed independently by resonant CW laser excitation.  Indeed, polariton bistability has been observed in the polariton ground state of this 0D system \cite{Paraiso2010}. Moreover, the photonic confinement is favorable to isolate the ground polariton level from intrinsic noise perturbation. This noise-free polariton state allows to accurately control the key parameters of the experiment, i.e. the extrinsic applied noise and the polariton bistability condition, which in turn, determine the characteristic time scale of the stochastic resonance. We report also the observation of stochastic resonance in 2D polariton gas in supplementary material. This extended study might stimulate the use of stochastic resonance in 2D polariton devices \cite{Ballarini2013, Anton2013,Gao2012}.\\
\begin{figure}
\includegraphics[scale=0.42]{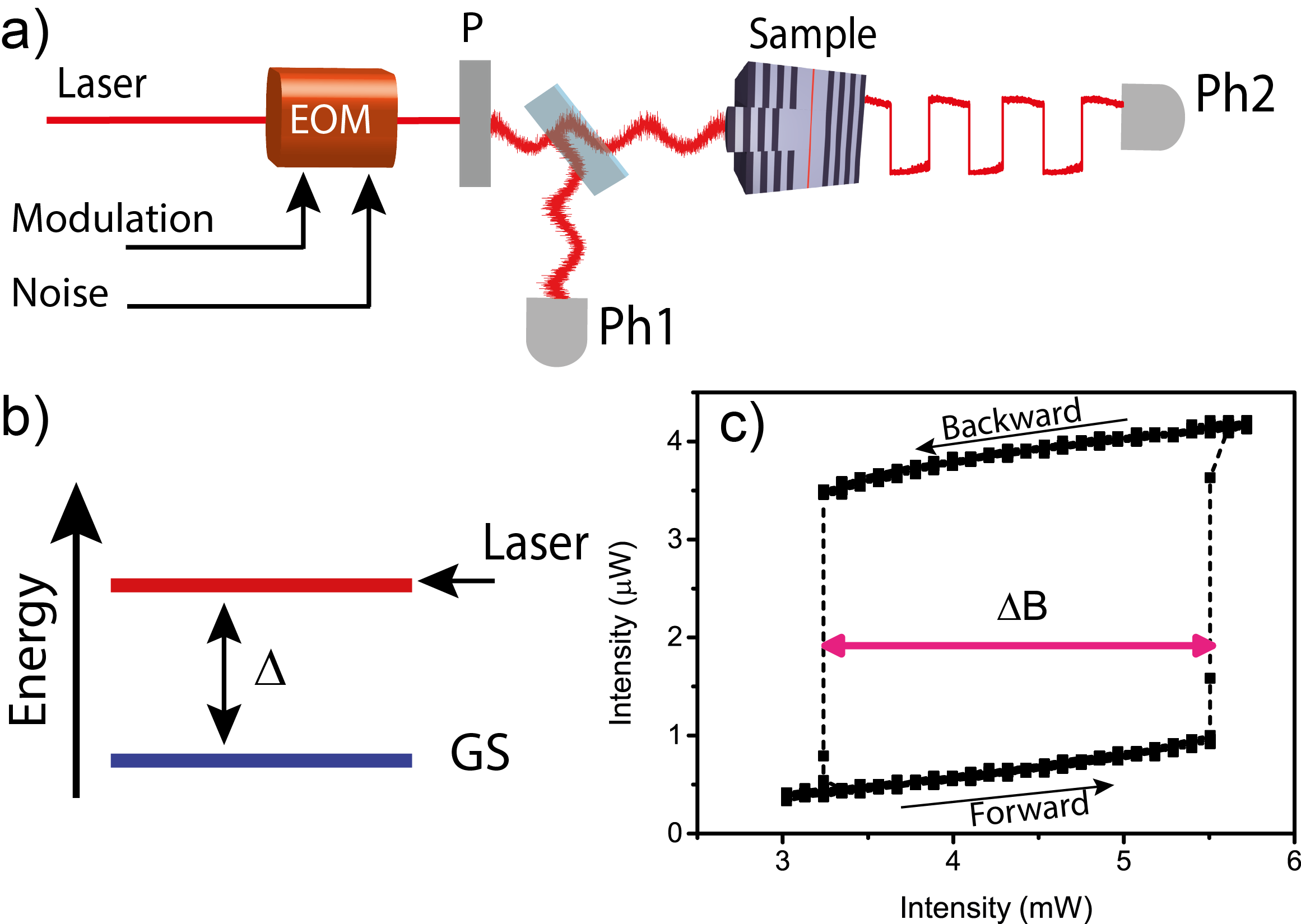}
\caption{a) Schematic of the experiment. A weak modulation is applied on the CW laser beam that excites resonantly a polariton gas on the lower state of the hysteresis. P, EOM and Ph$_{1,2}$ stand for the polarizer, the electro-optic modulator and fast photodiodes respectively. b) Energy diagram of the experimental configuration. GS is the polariton ground state. c) Polariton bistability obtained for a  negative exciton-cavity detuning (-5.12 meV) and a polariton-laser detuning of $\Delta=0.4$ meV. Bistability width is measured using the usual X-Y method \cite{Smith1977} that directly linked input and outut signals. The bistability width value is $\Delta\text{B}=2.27 \pm 0.43$ mW.}
\end{figure}
The microcavity sample is introduced in a cryostat and cooled down to 4 K. Experiments are performed at negative exciton-cavity detuning (-5.12 meV) for which the ground state polariton linewidth is in the order of $\gamma =100$ $\mu$eV. The sample is excited at normal incidence (k=0) with a continuous-wave mode Ti:Sapphire laser with 2$\%$ of noise intensity standard deviation. The polarization of the laser is prepared in a circular state to avoid any spinor related effects \cite{Paraiso2010}. Microcavity exciton-polaritons display large nonlinearities inherited from their excitonic counterpart that experience exciton-exciton interactions. Indeed, a spin polariton population will be energy blue shifted proportionally to its density. Therefore, this nonlinear system can be prepared in a bistable regime \cite{Baas2004} by blue detuned laser excitation respects to the polariton resonance. The bistability condition is fulfilled when $\Delta>\sqrt{3}\gamma$ where $\Delta$ stands for the energy detuning between the polariton mode and the laser frequency and $\gamma$ is the polariton linewidth. Figure 1c displays the polariton bistability for a detuning $\Delta=0.4 $meV.\\
To study the stochastic resonance, we fix the excitation laser power in the middle of the lower hysteresis state (Fig. 1c). By means of an electro-optic modulator, we imprint on the DC component power a 500 kHz bandwidth noise in addition of a weak periodic signal. The excitation is fully controllable and allows to vary the modulation amplitude and the frequency of the signal as well as the noise intensity standard deviation. It is then possible to study in detail the stochastic resonance response for a wide range of the input parameters. We record simultaneously the time evolution of the input excitation signal and the transmitted light of the polariton bistability by using 20 MHz bandwidth low noise photodiodes. Time streams are acquired through a 60 MHz oscilloscope bandwidth.\\
\begin{figure}
\includegraphics[scale=0.50]{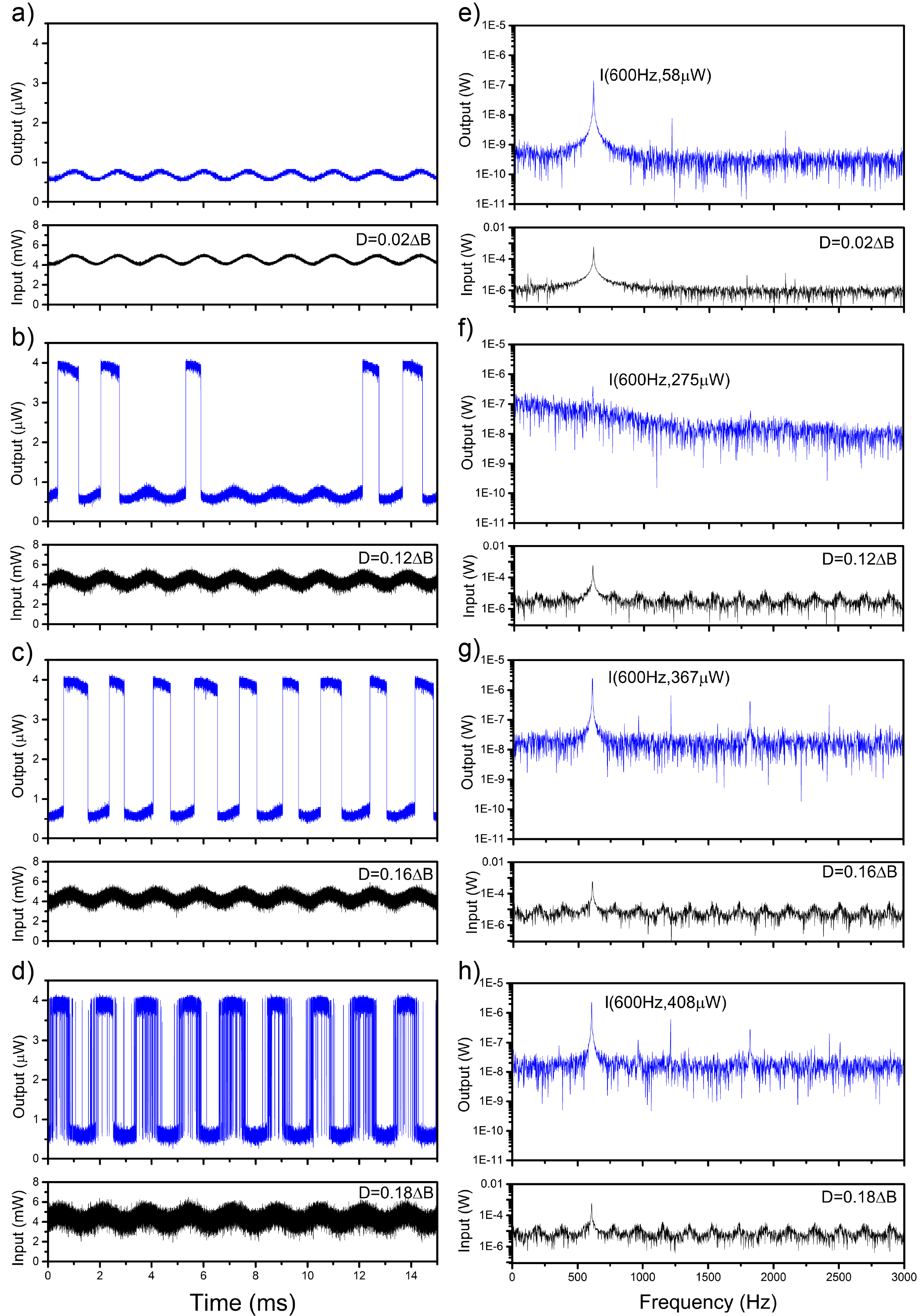}
\caption{Principle of the stochastic resonance. Time streams of the input (black) and output (blue) signals: From a) to d) the noise intensity is continuously boosted: 58$\mu$W (0.02$\Delta$B), 275 $\mu$W (0.12$\Delta$B), 367 $\mu$W (0.16$\Delta$B), 408 $\mu$W (0.18$\Delta$B). e) to h) show the corresponding frequency spectrum obtained by Fourier transforming time streams acquired for a 630 ms time window. The resolution of the spectrum is 1.5 Hz. Other parameters are the frequency of modulation $\nu=600$ Hz and the modulation amplitude of the input signal $A_0=525 \mu$W (0.23$\Delta$B). In c) and g), the weak transmitted periodic signal is coherently amplified by the cooperation of a noise perturbation through the polariton bistability.}
\end{figure}
For the purpose of investigating the stochastic resonance, we prepare the system with a polariton bistability width of $\Delta$B$=2.27 \pm 0.43$ mW and set the power signal at 4.5 mW in the middle of the bistable loop on the lower state (Fig. 1c). Then, by increasing the noise intensity standard deviation D, we record the transmitted signal in function of time. In Figure 2, we display the results recorded for a modulated signal at frequency $\nu_0=600$ Hz and amplitude $A_0= 525\mu$W corresponding to 0.23$\Delta$B. From top to bottom, we evidence the effect of the increased noise intensity on the time transmitted signal (Fig. 2a-d), and their corresponding frequency spectrum via Fourier transform (Fig. 2e-h). Residual peaks in the input signals at frequencies higher than 600 Hz are mainly due to the nonideal gaussian noise distribution provided by the electrical noise generator. Figure 2a shows the time transmitted signal when only the laser intensity fluctuations $D_0$= 58 $\mu$W (0.02$\Delta B$) are present. Note that the modulation amplitude is not sufficiently large to overcome the bistable intensity threshold by itself. The corresponding frequency spectrum (Fig. 2e) shows a weak peak amplitude, located at 600 Hz. By adding an external stochastic perturbation D= 275 $\mu$W (0.12$\Delta B$), the outcoming signal displays erratic jumps between the two stable states due to the random nature of the input signal (Fig. 2b, f). As we can see, the weak signal is almost buried by the noise background (Fig. 2f). Noise-induced-jumps start to display periodic hopping which tend to synchronize with the weak driving modulation frequency. Intriguingly, for an optimal noise intensity (Fig. 2c, g), the polariton bistability shows a coherent amplification of the weak transmitted signal due to the noise contribution D=367 $\mu$W (0.16$\Delta B$). The noise contributes coherently to switch the polariton gas between the two stable states. Notice that the spectrum displays combination of different harmonics. The Fourier series of a perfect square signal is composed of odd harmonics of the fundamental 600 Hz component. Here the spectrum also shows even harmonics. We attribute the appearance of those peaks to the stochastic origin of the resonance that imply a breathing of the output signal period. By introducing a larger amount of noise D=408 $\mu$W (0.18$\Delta B$), the transmitted signal shows again erratic jumps (Fig. 2d-h). For even larger noise intensity, the output signal is completely buried in the noise. The above description carries all the characteristics of the so-called "stochastic resonance".\\
To study the behavior of the polariton stochastic resonance in more detail, we plot a quantifier namely the spectral magnification factor :
\begin{equation}
M=\frac{I_{out}(\nu_0,D)}{I_{out}(\nu_0,D_0)} 
\end{equation}
\begin{figure}
\includegraphics[scale=1]{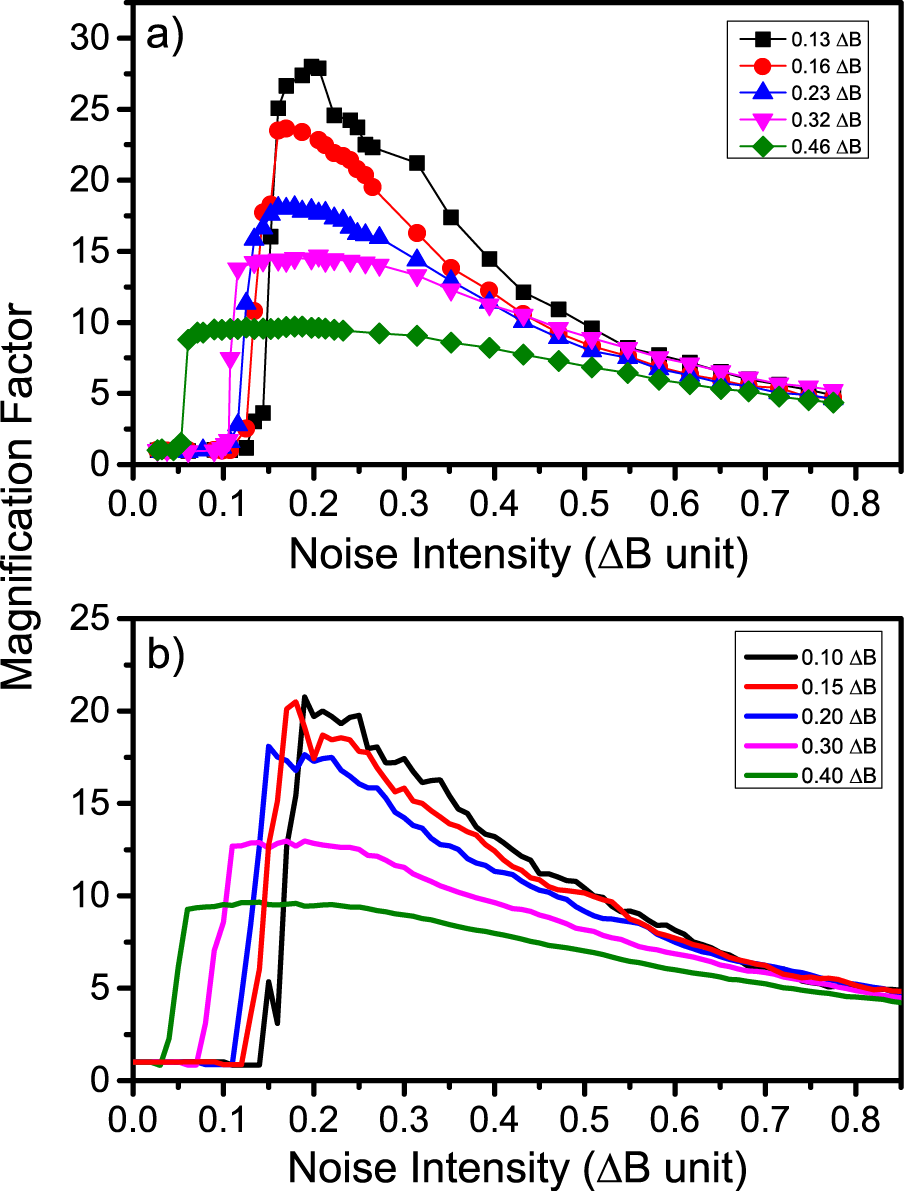}
\caption{a) Experimental and b) numerical spectral magnification factors for different amplitudes of modulation $A_0$. Numerical parameters are $\alpha_1=0.35$ meV, $\gamma=\hbar/10$ ps, $\Delta=0.45$ meV, $\nu_0=30$Hz. $A_0$ ranges from 0.1 to 0.4 $\Delta$B and the noise standard deviation varies from 0 to 0.8 $\Delta$B. $\Delta$B is the polariton bistability width.}
\end{figure}
Spectral peak intensities of the transmitted signal at the modulation frequency $\nu_0$ are extracted for different noise intensities $D$.  $I_{out}(\nu_0,D)$ spectral peak intensities are normalized by the transmitted signal intensity when only laser fluctuations exist $I_{out}(\nu_0,D_0)$.\\
We study the spectral magnification factor for a large range of noise intensities and modulation amplitudes from 0.29 mW (0.13 $\Delta$B) to 1.04 mW (0.46 $\Delta$B). The spectral magnification factor is plotted in Figure 3a for five amplitudes of modulation.\\
We clearly see a resonance behavior in function of the noise intensity. For a small modulation amplitude (0.13 $\Delta$B), the resonance appears for $D_{\text{SR}}\approx 0.2  \Delta$B, the weak transmitted signal shows a strong amplification. As the modulation amplitude increases, the stochastic resonance shifts to lower noise intensity accompagnied by a decreasing of the spectral peak intensity. This behavior reflects the tendency of the system to evolve to deterministic jumps between the two states without the cooperative interplay between noise and the weak signal. This effect evidences that, contrary to intuition, the weaker the modulated signal, the better can be the benefit of noise to magnify the signal transmission when being at the resonance operation point ($D\approx D_{\text{SR}}$).\\
The simplest approach used to model the stochastic resonance is indeed based on a two states model considering the Brownian motion of a particle in a double well potential \cite{Gammaitoni1998}. However, this model considers two discrete states. This reduces the dynamics only to the switching mechanism between the two states omitting all short time dynamics within the states. The dissipative nature of polariton bistability brings an asymmetry in the hysteresis respect to a symmetric bistability observed in a Schmitt-Trigger \cite{Fauve1983}. Therefore, we performed numerical simulations considering a stochastic perturbation of the Gross-Pitaevskii equation usually used to model polariton dynamics in bistability regime:
\begin{equation}
i \frac{d\Psi}{dt}=-\Delta\Psi-i\frac{\gamma}{2}\Psi+ \alpha_1|\Psi|^2\Psi+F
\label{GPE}
\end{equation}
where $\Psi$ and $\alpha_1$ are the polariton field and the interaction constant respectively. The present investigated stochastic resonance effect in polariton bistability fulfills the well established adiabatic approximation for which the driving field modulation frequency $\nu_0$ is orders of magnitude smaller than the intrinsic frequency dynamics of the polariton gas ($\gamma\approx$25 GHz) \citep{Gammaitoni1998}. Indeed, the CW resonant excitation of the polaritons ensures our working in the stationary regime. Since the driving field amplitude variation is extremely slow ($\mu$s time scale) respect to the transient response time of the polariton gas (ps time scale), it can be considered as a constant in equation \eqref{GPE}. Therefore, the polariton gas experiences a slow variation of the driving field at the noise correlation time scale:
\begin{equation}
F(t')=E_0+A_0  \cos(2\pi\nu_0t^{\prime}+\phi)+D(t^{\prime}),
\end{equation}
where $E_0$ is the DC amplitude component, $A_0$ the modulation amplitude  and $D(t^{\prime})$ the random perturbation term. The modulation frequency $\nu_0$ is set to 30 Hz. $\Delta$B is the bistability width obtained with the following set of parameters: $\alpha_1=0.35$ meV, $\gamma=\hbar/10$ ps, $\Delta=0.45$ meV with $\hbar$ the reduced Planck constant. The numerical correlation time of the Gaussian noise distribution is fixed to be $20 \mu$s. By Fourier transformation of simulated output time streams, the spectral magnification factor is calculated for five values of $A_0$ (Fig. 3b). Our numerical simulations reproduce qualitatively very well the experimental results of Figure 3a. We show a good agreement in terms of the noise activation intensities which shift to lower D when increasing the amplitude of the modulation. Moreover, the resonance peak and the overall shape of the spectral magnification factor are well reproduced by our simulation. The normalization of our results in terms of bistability width allow us to show a good agreement with works related to stochastic resonance in other physical systems where the activation noise intensity is expected to be $D_{\text{SR}}\approx0.2\Delta$B \cite{Gammaitoni1998}.\\
\begin{figure}
\includegraphics[scale=0.9]{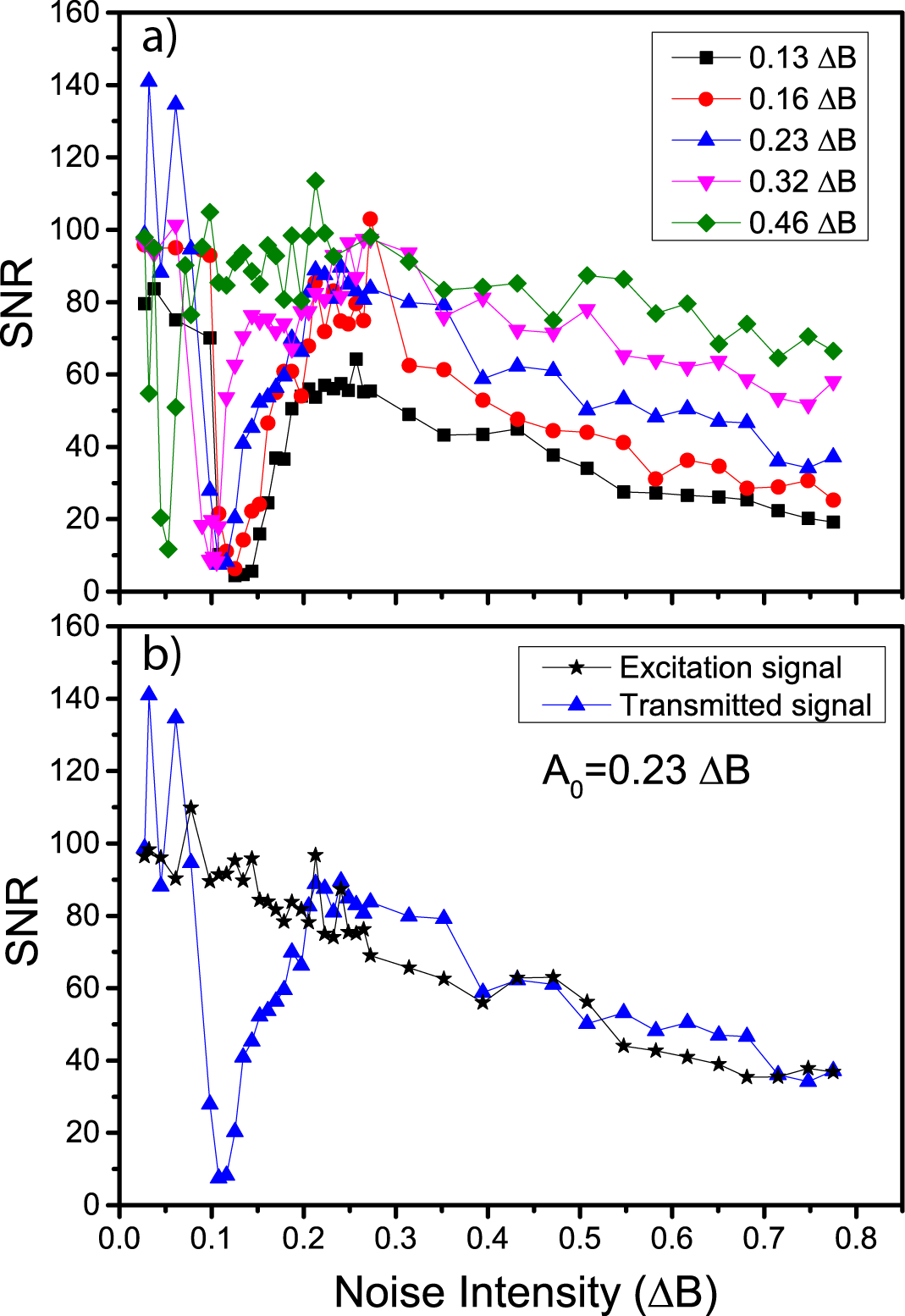}
\caption{a) Experimental signal-to-noise ratio for different modulation intensities. Experimental parameters are the same as in Figure 3a. b) Experimental SNR of the excitation signal (black) and corresponding transmitted signal (blue) for $A_0=0.23 \Delta$B in function of the noise standard deviation.}
\end{figure}
We plot in Figure 4 the signal-to-noise ratio defined as,
\begin{equation}
SNR=\frac{I_{out}(\nu_0,D)}{N(\nu,D)},
\label{SNR}
\end{equation}\\
where $N(\nu,D)$ is the noise background averaged between 650 Hz to 737 Hz. The behavior of the SNR differs from the previously plotted magnification factor. Indeed, the SNR peak value increases with the modulation signal amplitude (Fig. 4a) contrary to the decreasing behavior of the magnification peak value observed for the activation noise intensity $D_{\text{SR}}$. However, it is important to note that both trends are related to the same effect evoked in the paragraph above, namely the deterministic behavior of the system for amplitude of modulation close to the bistability width $\Delta$B. This behavior is in agreement with previously reported works \cite{Gammaitoni1989}.\\
In addition, we plot in Figure 4b the signal-to-noise ratio of the input and transmitted signals for modulation amplitude of $A_0=0.23\Delta$B. The input signal-to-noise ratio behavior follows an expected decrease inversely proportional to increasing noise intensity $N(\nu,D)$ in eq.\eqref{SNR}. Both SNRs show large values for very low noise contribution ($D<0.1\Delta$B).  This behavior, in the transmitted signal, is due to the dissipative character of polariton gas. Contrary to Schmitt-trigger switches \cite{Fauve1983}, the polariton bistability does not exhibit flat states. Actually the output polariton intensity varies linearly on both stable states in function of the pump power (Fig. 1c).  Moreover, the SNR of the transmitted signal shows a revival of the transmission at the D$_\text{SR}$ evidencing the stochastic resonance.\\
To conclude, we have reported observation of stochastic resonance in microcavity polaritons under bistable conditions. The experimental observation are well reproduced by a model based on the Gross-Pitaevskii equation. The observed behavior shows the potential of using stochastic resonance as a means to improve the coherent processing of a signal lost in noise. By this study, we highlight the opportunity of using noise as a control parameter to investigate polariton related phenomena. This first report opens appealing perspectives for the use of stochastic resonance as a tool to drive coherent polariton fluids either for application in all-optical processing or on fundamental aspects.
\\\\
We acknowledge fruitful discussions with Prof. Michiel Wouters and Prof. Iacopo Carusotto, and Philippe Cuanillon for technical support.\\
The present work has been supported by the Swiss National Science Foundation under project N°135003, the Quantum Photonics National Center of Competence in Research N°115509 and by the European Research Council under project Polaritonics contract N°219120. The polatom network is also acknowledged.




%
%

\end{document}